\documentclass[letterpaper,nobalancelastpage,raggedbottom,prb,twocolumn]{revtex4}
\usepackage{amsmath,amssymb,graphicx,times,setspace}

\def\Pc{{\phi_c}} 
\def\Pp{{\phi_p}}
 
\newcommand{\RM }[1]{\mathrm{#1}}
 
\newcommand{\pd }[2]{{\frac{\partial {#1}}{\partial {#2} }}}

\def\kB{ k_{\RM{B}} } 
\def\kv{\RM{K}}
\def\gcm{\RM{\,g/cm^3}} 
\def\ekt{{\epsilon/\kB T}}
\def\s2{-s_2/\kB} 
\newcommand{\sTwo}[1]{{-s_2({#1})/\kB}}
\def\Is2{I_{s_2}(r)}

\begin{document}
\title{Structural anomalies of fluids: \\
Origins in second and higher
  coordination shells} 
\author{William P. Krekelberg}
\email{krekel@che.utexas.edu} 
\affiliation{Department of Chemical
  Engineering, University of
  Texas at Austin, Austin, TX 78712.}  
\author{Jeetain Mittal}
\email{jeetain@helix.nih.gov} 
\affiliation{Laboratory of Chemical
  Physics, NIDDK, National Institutes of Health, Bethesda, MD
  20892-0520. }  
\author{Venkat Ganesan} 
\email{venkat@che.utexas.edu}
\affiliation{Department of Chemical Engineering and Institute for
  Theoretical Chemistry, University of Texas at Austin, Austin, TX
  78712.}  
\author{Thomas M. Truskett} 
\email{truskett@che.utexas.edu}
\thanks{Corresponding Author} 
\affiliation{Department of Chemical
  Engineering and Institute for Theoretical Chemistry, University of
  Texas at Austin, Austin, TX 78712.}

\begin{abstract}
  Compressing or cooling a fluid typically enhances its static
  interparticle correlations.  However, there are notable exceptions.
  Isothermal compression can reduce the translational order of fluids
  that exhibit anomalous waterlike trends in their thermodynamic and
  transport properties, while isochoric cooling (or strengthening of
  attractive interactions) can have a similar effect on fluids of
  particles with short-range attractions.  Recent simulation studies
  by Yan {\em et al.} [Phys. Rev. E {\bf 76}, 051201 (2007)] on the
  former type of system and Krekelberg {\em et al.} [J. Chem. Phys.
  {\bf 127}, 044502 (2007)] on the latter provide examples where such
  structural anomalies can be related to specific changes in second
  and more distant coordination shells of the radial distribution
  function.  Here, we confirm the generality of this microscopic
  picture through analysis, via molecular simulation and integral
  equation theory, of coordination shell contributions to the two-body
  excess entropy for several related model fluids which incorporate
  different levels of molecular resolution.  The results suggest that
  integral equation theory can be an effective and computationally
  inexpensive first-pass tool for assessing, based on the pair
  potential alone, whether new model systems are good candidates for
  exhibiting structural (and hence thermodynamic and transport)
  anomalies.
\end{abstract}

\maketitle

\section{Introduction}
\label{sec:intro}

A bulk equilibrium fluid is translationally invariant; i.e., its
one-particle density, $\rho^{(1)}({\bf r})=\rho$, is constant.
Nonetheless, assuming spherically-symmetric interactions, the local
density $\rho g(r)$ surrounding a reference particle is a function of
distance~$r$ from its center, where $g(r)$ is the radial distribution
function (RDF) of the fluid.\cite{Hansen2006Theory-of-Simpl} Although
the RDF depends on both the form of the interparticle interactions and
the thermodynamic state, some features of its shape are fairly
general.  For example, the RDF vanishes for $r$ less than the
effective exclusion diameter of the particles.  For larger $r$, it
shows an oscillatory decay toward unity with peaks loosely
corresponding to coordination ``shells''.  Away from the critical
point, the structure of the RDF typically persists for distances
comparable to a few particle diameters, reflecting the short range of
the interparticle correlations.

Studies of the liquid state have primarily focused on the particles in
the first coordination shell.  This is due in part to the important
role that nearest neighbors are expected to play in determining many
physicochemical properties.  For example, both the non-ideal
contribution to the equation of state of the hard-sphere fluid (see,
e.g., Ref.~\onlinecite{Lowen2000Fun-With-Hard-S}) and the collision
frequency in Enskog theories for transport
processes\cite{Chapman1970The-Mathematica} scale with the ``contact''
density $\rho g(\sigma)$, where $\sigma$ is the particle diameter.
The hard-sphere equation of state is the standard reference system for
perturbation theories.\cite{Hansen2006Theory-of-Simpl} It also
accurately predicts how the thermodynamics of ``hard-sphere''
colloidal suspensions relate to their structure, as has recently been
experimentally verified by confocal
microscopy.\cite{Dullens2006Direct-measurem} Furthermore, analysis of
first-shell contributions to hydration structure and thermodynamics
helps to understand and make predictions about a wide variety of
aqueous solution
properties.\cite{Nemethy1962Structure-of-Wa,Pratt1985Theory-of-Hydro,Beglov1994Finite-represen,Matubayasi1994Thermodynamics-,Silverstein1999Molecular-model,Lockwood1999Evaluation-of-F,Lockwood2000Functional-Grou,Hummer2000New-perspective,Paulaitis2002Hydration-theor,Asthagiri2004Hydration-Struc,Dill2005MODELING-WATER-}

Although the second shell of the RDF has received comparatively less
attention, there is evidence that it also contains structural
information relevant for understanding nontrivial behaviors of
liquids.  One notable feature is the
shoulder\cite{Truskett1998Structural-prec} that it develops near the
freezing transition, which in turn becomes a pronounced split
peak\cite{Finney1970Random-Packings,Bennett1972Serially-Deposi,Rahman1976Molecular-dynam,Stillinger1984Packing-Structu,Hiwatari1984Structural-char,Clarke1987Numerical-simul,Tsumuraya1990Local-structure,Snook1991Structure-of-co,Donev2005Pair-correlatio,OMalley2005Structure-of-ha}
in supercooled liquid and glassy states.  Analysis of the
configurations that give rise to this structural motif indicate that
it reflects frustration of
icosahedral\cite{Tsumuraya1990Local-structure} and emerging
crystalline\cite{Truskett1998Structural-prec,OMalley2005Structure-of-ha}
order in the fluid.  Understanding how these these types of structural
features connect to relaxation processes of supercooled liquids is an
active area of research (see, e.g.,
Refs.~\onlinecite{Dzugutov2002Decoupling-of-D,Jain2005Role-of-local-s,Shintani2006Frustration-on-,Kawasaki2007Correlation-bet}).

In this work, however, we focus on the second and higher coordination
shells of the RDF for a different reason: to understand their role in
the {\em structural anomalies} of fluids.  Interparticle correlations
of most fluids are enhanced upon (i) compression or (ii) cooling
(alternatively, strengthening of interparticle attractions).
Nonetheless, there are a few systems of scientific interest that
exhibit notably different behaviors.  For example, compression induced
disordering occurs in water and other fluids with anomalous waterlike
trends in their thermodynamic and transport
properties.\cite{Errington2001Relationship-be,Shell2002Molecular-struc,Truskett2002A-Simple-Statis,Yan2005Structural-Orde,Yan2006Family-of-tunab,Oliveira2006Structural-anom,Mittal2006Quantitative-Li,Mittal2006Relationship-be,Sharma2006Entropy-diffusi,Oliveira2007Interplay-betwe,Agarwal2007Ionic-melts-wit,Agarwal2007Waterlike-Struc,Yan2007Structure-of-th,Oliveira2007in-press}
Cooling (or attraction) induced disordering, on the other hand, can
occur in fluids of particles with short-range attractive (SRA)
interactions,\cite{Mittal2006Quantitative-Li,Krekelberg2007How-short-range}
such as concentrated suspensions of colloids.

These anomalies do not appear to be first-shell effects.  Rather, they
reflect how structuring in second and more distant coordination shells
responds to changes in thermodynamic or system parameters.  For
example, Yan {\em{et al.}}\cite{Yan2007Structure-of-th} recently
demonstrated in an insightful paper how the structural anomaly of the
five-site transferable interaction potential (TIP5P) model\cite{Mahoney2000A-five-site-mod} for
water is quantitatively related to
compression induced translational disordering of molecules in the
second coordination shell.  Similarly, Krekelberg {\em{et
    al.}}\cite{Krekelberg2007How-short-range} have shown that the
cooling (or attraction) induced structural anomaly of a square-well
SRA fluid is due to weakening of second- and higher-shell pair
correlations.

The goal here is to study the generality of the above findings.  It is
known that a number of models, with varying levels of molecular
resolution, can qualitatively predict the structural anomalies of the
aforementioned
systems.~\cite{Errington2001Relationship-be,Oliveira2006Structural-anom,Yan2007Structure-of-th,Mittal2006Quantitative-Li,Mittal2006Relationship-be,Krekelberg2007How-short-range,Shell2002Molecular-struc,Truskett2002A-Simple-Statis,Oliveira2007Interplay-betwe,Agarwal2007Ionic-melts-wit,Sharma2006Entropy-diffusi,Agarwal2007Waterlike-Struc,Oliveira2007in-press,Yan2005Structural-Orde,Yan2006Family-of-tunab}
But do the anomalies exhibited by lower resolution models have the
same microscopic origins as those of more detailed models?  Moreover,
can the behavior of the lower resolution models be predicted, at least
qualitatively, by integral equation theory?  If so, it would suggest
that integral equation theory might serve as a valuable first-pass
tool in assessing, based on the pair potential alone, whether new
model systems might be good candidates for exhibiting structural
anomalies.

Furthermore, although the structurally anomalous trends analyzed here
are interesting in their own right, there is a more compelling reason
to try to understand their origins.  In short, they appear to be
closely linked to other distinctive dynamic and thermodynamic
behaviors.  For example, in addition to being ``structurally
anomalous'', cold liquid water is also ``dynamically anomalous'' in
that its self-diffusivity increases upon isothermal compression and
``thermodynamically anomalous'' in that its volume increases upon
isobaric cooling.  Errington and Debenedetti\cite{Errington2001Relationship-be} first noticed that these
particular anomalies form a cascade in the temperature-density plane
for the extended simple point charge (SPC/E) model\cite{Berendsen1987The-Missing-Ter} of
water.\cite{Errington2001Relationship-be} Specifically, the
thermodynamic anomaly occurs only for state points that also exhibit
the dynamic anomaly.  The dynamic anomaly, in turn, is only present
for states that also exhibit the structural anomaly.  Strong
correlations between these three basic types of anomalies have since
been documented for a wide variety of model systems with waterlike
properties.\cite{Oliveira2006Structural-anom,Yan2007Structure-of-th,Shell2002Molecular-struc,Truskett2002A-Simple-Statis,Oliveira2007Interplay-betwe,Sharma2006Entropy-diffusi,Oliveira2007in-press,Kumar2005Static-and-dyna,Esposito2006Entropy-based-m,Netz2006Thermodynamic-a,Xu2006Thermodynamics-,Szortyka2007Diffusion-anoma}

A similar connection between structural and dynamic anomalies has now
also been identified for model SRA
fluids.\cite{Mittal2006Quantitative-Li,Krekelberg2007How-short-range}
In those systems, the most commonly studied dynamic anomaly is an {\em
  increase} in self-diffusivity upon cooling (or strengthening of
interparticle attractions), which can occur at sufficiently high
particle
concentrations.\cite{Eckert2002Re-entrant-Glas,Pham2002Multiple-Glassy,Bergenholtz1999Nonergodicity-t,Fabbian1999Ideal-glass-gla,Dawson2001Higher-order-gl,Zaccarelli2002Confirmation-of,Sciortino2002One-liquid-two-}
Krekelberg {\em et al.}\cite{Krekelberg2007How-short-range} discovered
that the self-diffusivity anomaly for a square-well SRA fluid occurs
only for state points that also exhibit the cooling (or attraction)
induced structural anomaly discussed above.  In other words, it
appears that SRA fluids can also display a cascade of anomalies
qualitatively similar to those of waterlike fluids.

Although structural and dynamic properties of these systems show
unusual dependencies on quantities like temperature or density, the
correlations between structure and dynamics are often similar to those
found in simpler liquids (e.g., the hard-sphere
fluid).\cite{Mittal2006Quantitative-Li,Mittal2006Relationship-be,Errington2006Excess-entropy-,Sharma2006Entropy-diffusi,Krekelberg2007How-short-range}
In fact, it was recently
demonstrated\cite{Errington2006Excess-entropy-} that the cascade of
anomalies of one waterlike model system can be semi-quantitatively
predicted based only on knowledge of the state dependencies of excess
entropy, which measures structural
order,\cite{Truskett2000Towards-a-quant} and quasi-universal excess
entropy
scalings\cite{Rosenfeld1999A-quasi-univers,Rosenfeld1977Relation-betwee,Dzugutov1996A-univeral-scal}
for the transport coefficients.  All of this suggests that
investigations like the present one, which probe the physics of
structural anomalies, might also provide insights into dynamic and
thermodynamic anomalies as well.

\section{Methods}
\label{sec:sim_methonds}

We used molecular dynamics simulation and integral equation theory to
examine various models from two classes of fluids known to exhibit
structural anomalies: those with waterlike properties and those
comprising particles with SRA interactions. For the integral equation
theory analysis, we numerically solved the Ornstein-Zernike
equation\cite{Ornstein1914integral-equati} together with an
approximate closure relation using the method of Labik \textit{et
  al.}\cite{Labik1985A-RAPIDLY-CONVE} In the discussion of the models
below, we mention the specific closures employed and provide further
details about the molecular simulations.

We did not perform a systematic study here to determine which of many
possible closure relations\cite{Hansen2006Theory-of-Simpl} provides the most quantitatively accurate 
description for each model.  Rather, our focus 
was to explore whether integral
equation theory solved with standard closure relations,
such as Percus-Yevick (PY)\cite{Percus1958Analysis-of-Cla} or
hypernetted-chain (HNC)\cite{van-Leeuwen1959New-method-for-}, can in fact
qualitatively predict both the structural anomalies and their
microscopic origins in the RDF.  Molecular simulations of the model
systems provide the data necessary to make that basic 
determination.

\subsection{Waterlike fluid models}

We investigated two waterlike models: (1) the SPC/E
\cite{Berendsen1987The-Missing-Ter} model and (2) a lower resolution
``core-softened''
\cite{Oliveira2006Thermodynamic-a,Oliveira2006Structural-anom} model.
We chose the SPC/E model because it represents one of the most
commonly studied effective pair potentials for water, and it is known
to qualitatively reproduce many of water's distinctive thermodynamic,
dynamic, and structural properties.\cite{Errington2001Relationship-be}
As a result, it provides a reasonable baseline against which to
compare simpler, lower resolution models.  Details of our molecular
dynamics simulations of the SPC/E model are the same as reported in
Ref.~\onlinecite{Mittal2006Quantitative-Li}.

\begin{figure}[htbp]
  \centering
  \includegraphics[keepaspectratio]{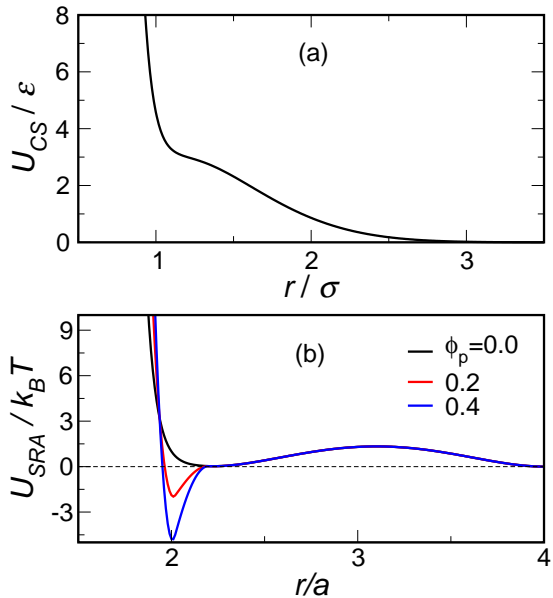}
  \caption{(a) Pair potential of the core-softened model $U_{\RM{CS}}(r/\sigma)/\epsilon$
    [see Eq.~\eqref{eq:core_softened_potential}].  (b) Pair potential
    of the model SRA fluid $U_{\RM{SRA}}(r/a)/\kB T$ discussed in the
    text for various values of polymer concentration $\Pp$.  Further
    details on this SRA model are provided in
    Refs.~\onlinecite{Puertas2002Comparative-Sim} and
    \onlinecite{Puertas2003Simulation-stud}.}
  \label{fig:CS_potential}
\end{figure}

The core-softened
model\cite{Oliveira2006Thermodynamic-a,Oliveira2006Structural-anom}
that we studied is more schematic.  It is defined by the effective
pair potential $U_{\RM{CS}}(r)$ [see
Figure~\ref{fig:CS_potential}(a)],
\begin{equation}
  \label{eq:core_softened_potential}
  U_{\RM{CS}}(r)=4\epsilon\Biggl[\biggl(\frac{\sigma}{r}\biggr)^{12}-\biggl(\frac{\sigma}{r}\biggr)^6\Biggr]
  +5\epsilon\RM{exp}\Biggl[-\biggl(\left[\frac{r}{\sigma}\right]-0.7\biggr)^2\Biggr],
\end{equation}
where $\epsilon$ is the characteristic energy scale.  The main idea
behind this potential is that it has two different kinds of repulsions
that act at different length scales.  The harsh $(\sigma/r)^{12}$
repulsion defines the effective hard-core diameter ($\sigma$), while
the softer Gaussian repulsion extends to considerably larger
distances.  The end result is that the average interparticle
separation, and hence the density, of this fluid can depend
sensitively on both temperature and pressure.  The model is similar to
cold water in that it favors locally open (low-density) structures at
moderate pressure and low temperature, but can collapse to denser
structures when compressed or heated enough to overcome the soft
Gaussian repulsion.  Although this low resolution model does not
provide an accurate molecular-level description of water, it does
qualitatively reproduce many of its peculiar thermodynamic,
structural, and kinetic
behaviors.\cite{Oliveira2006Thermodynamic-a,Oliveira2006Structural-anom,Mittal2006Relationship-be,Oliveira2007Interplay-betwe}

To compute the properties of the core-softened model, we performed
molecular dynamics simulations in the microcanonical ensemble using
$N=1000$ identical particles of mass $m$.  We used the velocity-Verlet
technique for integrating the equations of motion with a time step of
$\Delta t=0.002\sigma\sqrt{m/\epsilon}$.  For the integral equation
theory analysis, we employed the HNC
closure.  We chose the HNC approximation because of its ability
to describe the structure of another fluid with a soft Gaussian repulsion,
the Gaussian-core model.\cite{Louis2000Mean-field-flui}
We investigated both
the SPC/E and core-softened models over a wide range of density and
temperature, where they are known to exhibit structural
anomalies.\cite{Errington2001Relationship-be,Oliveira2006Thermodynamic-a,Oliveira2006Structural-anom,Mittal2006Relationship-be,Oliveira2007Interplay-betwe}

\subsection{SRA fluid models}
\label{sra_models}

The first SRA fluid model that we considered qualitatively describes a
solution of (explicit) colloidal particles attracted to one another by
depletion interactions due to the presence of (implicit) non-adsorbing
polymers. The details of the colloidal pair potential are provided in
Refs.~\onlinecite{Puertas2002Comparative-Sim} and
\onlinecite{Puertas2003Simulation-stud}, but we discuss some of its
main features below.  The colloids are spherical and their effective
interactions consist of three parts.  The first is a steeply
repulsive, essentially hard-sphere (HS), contribution $U_{\mathrm{HS}}
(r) = \kB T (2 a/r)^{36}$, where $2a$ is the colloid diameter, $\kB$
is Boltzmann's constant, and $T$ is temperature.  The second term
represents the aforementioned polymer-induced depletion attraction
$U_{\RM{AO}} (r)$, approximated by the
Asakura-Oosawa\cite{Asakura1958depletion} potential.  The strength of
this attraction is proportional to the volume fraction of polymers in
solution $\Pp$, while the range is controlled by the radius of
gyration of the polymers $R_{\mathrm g}$, set in this case to $a/5$.
A soft repulsion $U_{\RM{R}}$ is also added to the effective
interparticle potential\cite{Puertas2002Comparative-Sim} to prevent
fluid-fluid phase separation.  Figure~\ref{fig:CS_potential}(b)
displays the total colloidal potential
$U_{\RM{SRA}}=U_{\RM{HS}}+U_{\RM{AO}}+U_{\RM{R}}$ for three different
polymer concentrations $\Pp$. The details of the molecular dynamics
simulations that we performed for this fluid are the same as those
reported in Ref.~\onlinecite{Krekelberg2006Free-Volumes-an}, with one
exception.  In the original study, a weakly polydisperse system was
investigated.  Here, all particles considered had identical radius $a$
and mass $m$.  The advantage of focusing on a monodisperse system is
that the pair correlations are unambiguously described by a single
RDF, which facilitates the analysis discussed in the next section.
For the integral equation theory of this SRA fluid, we employed the
PY closure.  The PY approximation is a natural choice here due to its
simplicity and its ability to describe the structure of liquids
with harshly repulsive, short-range
potentials,\cite{Hansen2006Theory-of-Simpl} (in particular, 
other SRA fluids\cite{Dawson2001Higher-order-gl}). 

We also considered a simpler model SRA fluid: a system of identical
square-well particles with attractive well depth -$\epsilon$ and width
0.03$\sigma$, where $\sigma$ represents the hard-core diameter.  This
model is similar to others known to exhibit
structural\cite{Krekelberg2007How-short-range} and dynamic
\cite{Krekelberg2007How-short-range,Zaccarelli2002Confirmation-of}
anomalies. We also use the PY closure in our integral equation theory
analysis of this fluid for the reasons mentioned above.

\subsection{Quantification of structural order}

For each of the model fluids, we calculated the state dependencies of 
$-s_2/\kB$,
\begin{equation}
  \label{eq:s2}
  -\frac{s_2}{\kB}=2\pi\rho \int_0^\infty r^2 \{ g(r) \RM{ln}\, g(r) -[g(r)-1]\},  
\end{equation}
where $s_2$ is the translational pair-correlation
contribution\cite{Nettleton1958Expression-in-T,Baranyai1989Direct-entropy-}
to the excess entropy and $\rho$ is the number density.  We used the
orientationally averaged oxygen-oxygen RDF in Eq.~\ref{eq:s2} for the
analysis of SPC/E water.  It has been shown that $-s_2/\kB$ not only
quantifies the translational order exhibited by a fluid (the tendency
of pairs of particles to adapt preferential
separations),\cite{Truskett2000Towards-a-quant} but it also strongly
correlates with the transport coefficients (see, e.g., Refs.
\onlinecite{Dzugutov1996A-univeral-scal,Mittal2006Quantitative-Li,Krekelberg2007How-short-range}).
Other translational order parameters have also been introduced to
study the structure of molecular and colloidal
fluids,\cite{Truskett2000Towards-a-quant,Torquato2000Is-Random-Close,Errington2001Relationship-be}
but these measures are known to correlate strongly with
$s_2$,\cite{Truskett2000Towards-a-quant,Errington2001Relationship-be}
and thus we exclusively use the latter in our analysis.

To understand how the various coordination shells of the RDF
contribute to $-s_2/\kB$, we also investigated the cumulative order integral
$\Is2$, defined as\cite{Krekelberg2007How-short-range}
\begin{equation}
  \label{eq:Is2}
  I_{s_2}(r)=2\pi\rho\int_0^r r'^2\{g(r')\RM{ln}g(r')-[g(r')-1]\}dr'.
\end{equation}
 Note that $I_{s_2}(r)\rightarrow
-s_2/\kB$ as $r\rightarrow \infty$.  

Finally, we adopted the following criteria to identify structurally
anomalous behavior.
\begin{subequations}
  \label{eq:anomalies}
\begin{align}
  \left(\pd{[-s_2]}{\rho}\right)_T&<0,&\text{$\rho-$structural
    anomalies}   \label{eq:anomaliesDens} \\
  \left(\pd{[-s_2]}{[\kB
      T/\epsilon]}\right)_\rho&>0,&\text{$T-$structural
    anomalies}\label{eq:anomaliesT}
\end{align}
\end{subequations}
As indicated in the Introduction, waterlike fluids exhibit
$\rho-$structural
anomalies\cite{Mittal2006Quantitative-Li,Errington2006Excess-entropy-}
and SRA fluids display $T-$structural
anomalies.\cite{Krekelberg2007How-short-range}

\section{Structural anomalies}

\subsection{Waterlike fluids}
\label{sec:struct-anom-waterl}

\subsubsection{SPC/E water}

\begin{figure}[htb]
  \centering \includegraphics{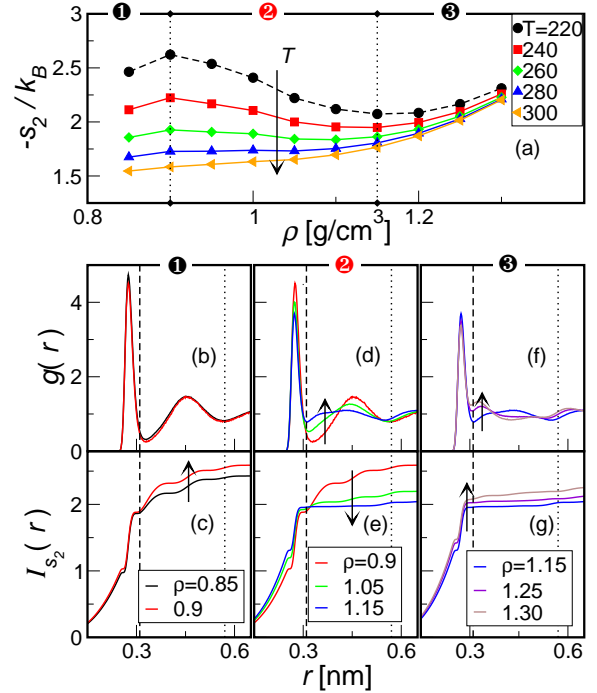}
  
  \caption{Structural data for the SPC/E water model obtained from
    molecular dynamics simulations.  (a) Structural order parameter
    $-s_2/\kB$ as a function of density $\rho$ at
    $T=220\kv,240\kv,260\kv,280\kv$ and $300\kv$. Vertical dotted
    lines are at $\rho=0.9 \gcm$ and $\rho=1.15 \gcm$, the approximate
    boundaries for the region of anomalous structural behavior. 
(Lower panel) Orientationally averaged oxygen-oxygen radial
    distribution function $g(r)$ and cumulative order integral
    $I_{s_2}(r)$ along the $T=220\kv$ isotherm [black circles, dashed
    curve in (a)] for three different density regions: (b,c) $\rho
    \leq 0.9\gcm$ [up to maximum in $-s_2(\rho)/\kB$], (d,e) $0.9\gcm
    \leq \rho \leq 1.15\gcm$ [between maximum and minimum in
    $-s_2(\rho)/\kB$], (f,g) $\rho \geq 1.15\gcm$ [beyond minimum in
    $-s_2(\rho)/\kB$].  The regions are indicated by circled numbers
    along top of (a) and lower panel.  In the lower panel, arrows indicate
    direction of increasing density; dashed vertical line is at
    $r=0.31\,\RM{nm}$ and dotted vertical line is at
    $r=0.57\,\RM{nm}$, the approximate locations of the first and
    second minima in $g(r)$, respectively.}.
  \label{fig:spce_s2fig}
\end{figure}

First, we discuss the simulation results for how $-s_2/\kB$ (i.e.,
translational order) of the SPC/E water model changes with density
$\rho$.  As can be seen in Figure~\ref{fig:spce_s2fig}(a), SPC/E water
displays the $\rho-$structural anomalies of
Eq.~\eqref{eq:anomaliesDens} over the density range $0.9\gcm \leq \rho
\leq 1.15\gcm$ and $T<280\kv$. To gain insights into the origins of
this behavior, we examine the orientationally averaged oxygen-oxygen
RDF and $I_{s_2}$ as a function of $\rho$ along the $T=220\kv$
isotherm for three different density regions: (1) the initial increase
of $\sTwo{\rho}$ at low densities [$\rho\leq 0.9 \gcm$,
Figs.~\ref{fig:spce_s2fig}(b,c)], (2) the anomalous decrease of
$\sTwo{\rho}$ at intermediate densities [$0.9 \gcm \leq \rho \leq 1.15
\gcm$, Figs.~\ref{fig:spce_s2fig}(d,e)], and (3) the increase of
$\sTwo{\rho}$ at high densities [$\rho>1.15 \gcm$,
Figs.~\ref{fig:spce_s2fig}(f,g)].

Compressing the fluid in the lower-density region (1) $(\rho\leq
0.9\gcm)$ has relatively little effect on the RDF
[Fig.~\ref{fig:spce_s2fig}(b)], but it does lead to a small net
increase in translational order.  As can be seen from the behavior of
$\Is2$ in Fig.~\ref{fig:spce_s2fig}(c), the changes come primarily
from the second shell.  The reason is that the coordination number
of water (approximately four, reflecting local tetrahedral
hydrogen-bonding to nearest neighbors) 
is insensitive to changes in density over
this range.\cite{Sciortino1991Effect-of-defec} As a result, the
increase of $\rho$ is compensated by a slight decrease in the first peak
of the RDF, and thus the first-shell contribution to the structural
order remains largely unchanged.  In the second shell, however, 
the change in density does not affect the RDF 
(i.e., the strength of the correlations with the central molecule).
This means that compression induced hydrogen-bond bending has allowed 
more total water molecules into the
second shell, which in turn leads to an overall increase in 
translational order.

On the other hand, 
further increases in density [region (2), $0.9\gcm \leq \rho \leq
1.15\gcm]$ result in a pronounced decrease in $-s_2(\rho)/\kB$, i.e.
the $\rho$-structural anomaly.  As can be seen in
Fig.~\ref{fig:spce_s2fig}(d), the main implications of compression for
the interparticle correlations are a dramatic flattening of the second
coordination shell and an associated shifting inward of these
molecules into the interstitial space between the first and second
shells. These structural changes are
consistent with the earlier simulation 
observation\cite{Sciortino1991Effect-of-defec} that high local
density can force a fifth molecule from the second shell into the
periphery of the otherwise four-coordinated first shell.  
Inspection of $\Is2$ [Fig.~\ref{fig:spce_s2fig}(e)] confirms
that the decrease in structural ordering is almost entirely due to
reduced correlations between the central and second-shell molecules.
In fact, Yan {\em et al.} convincingly demonstrated that a similar structural
anomaly in TIP5P water can also be attributed to compression induced
translational disordering of the second
shell.\cite{Yan2007Structure-of-th}  

Can these structural changes explain water's self-diffusity anomaly?
Sciortino {\em et al.}\cite{Sciortino1991Effect-of-defec} argued,
based on molecular simulation results, that the presence of a 
fifth molecule in the first
coordination shell significantly lowers 
the barriers for translational and rotational motions of the central 
water molecule.  This suggests that second-shell waters play a central
role in water's increased mobility under compression.      
Interestingly, since the self-diffusivity of SPC/E water is strongly
correlated to $s_2$ over these
conditions,\cite{Mittal2006Quantitative-Li} one can independently 
draw the same conclusion from the data in Fig.~\ref{fig:spce_s2fig}. 

Finally, we observe that, at sufficiently high densities [region (3),
$\rho>1.15\gcm$], translational order again increases upon
compression.  This is ``normal'' behavior for dense liquids, and it
simply reflects the fact that smaller volumes force particles to adopt
locally ordered (i.e., efficient) packing structures.\cite{Truskett2000Towards-a-quant,Errington2001Relationship-be}

\subsubsection{Core-softened model}

\begin{figure}[htb]
  \centering
  \includegraphics{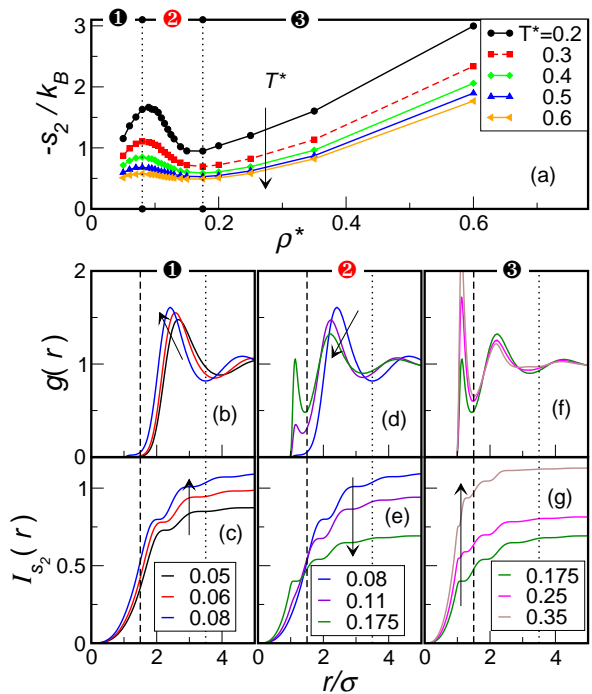}
  \caption{Structural data obtained from molecular dynamics simulations
    of the core-softened potential discussed in the text.  (a)
    Structural order parameter $-s_2/\kB$ as a function of reduced
    density $\rho^*=\rho \sigma^3$ at $T^*=\kB
    T/\epsilon=0.2,0.3,0.4,0.5$ and $0.6$, where $\sigma$ is the
    particle diameter, and $\epsilon$ is the energy scale of the
    potential (see Eq.~\eqref{eq:core_softened_potential}).  Arrow
    indicates direction of increasing $T^*$, and vertical dotted lines
    are at $\rho^*=0.08$ and $\rho^*=0.175$, the approximate
    boundaries of the region of anomalous structural behavior. (Lower
    panel) Radial distribution function $g(r)$ and cumulative order
    integral $I_{s_2}(r)$ along the $T^*=0.3$ isotherm [red squares,
    dashed curve in (a)] for three density regions: (b,c) $\rho^* \leq
    0.08$ [up to $-s_2(\rho^*)/\kB$ maximum], (d,e) $0.08 \leq \rho^*
    \leq 0.175$ [between maximum and minimum in $-s_2(\rho^*)/\kB$],
    (f,g) $\rho^*\geq 0.175$ [beyond minimum in $-s_2(\rho^*)/\kB$].
    The regions are indicated by circled numbers along top of (a) and
    lower panel. In lower panels, arrows indicate direction of
    increasing density; numbers in legends indicate values of
    $\rho^*$; vertical dashed line is at $r=1.5\sigma$ and vertical
    dotted line is at $r=3.5\sigma$, the approximate locations of the
    first and second minima in $g(r)$, respectively.}
  \label{fig:ramp_sim_s2fig}
\end{figure}

In this section, we investigate how the translational order of the
lower resolution, core-softened model of
Eq.~\ref{eq:core_softened_potential} responds to changes in density.
First, we consider the results from the molecular dynamics
simulations.  One striking feature of the data is that the behavior of
$-s_2/\kB$ as a function of reduced density $\rho^*=\rho\sigma^3$,
displayed in Fig.~\ref{fig:ramp_sim_s2fig}(a), is qualitatively
similar to that of SPC/E water [see Fig.~\ref{fig:spce_s2fig}(a)].
Specifically, the core-softened model also displays $\rho$-structural
anomalies over the density range $0.08 \leq \rho^* \leq 0.175$ that
become more pronounced at lower temperature.

Clearly, the core-softened model is very different from the SPC/E
model in that the former does not provide a molecular description of
water, and thus it does not favor the formation of tetrahedrally
coordinated hydrogen-bond networks, etc.  Nonetheless, as we explain
below, the main ``microscopic'' origins of its density-dependent
trends in structural order are basically the same as those for the
SPC/E model.

In order to appreciate the similarity between these two models, it is
helpful to first notice one difference.  In the SPC/E model, the
attractive ``hydrogen-bond'' interactions promote the formation of a
first coordination shell, even at relatively low density.  In
contrast, since there are no attractions in the core-softened model,
the Gaussian repulsion prevents the ``first'' coordination shell (near
the hard-core diameter, $1.0\leq r/\sigma\leq 1.5$) from forming until
sufficiently high density ($\rho^*\gtrsim 0.1$).  On the other hand,
the ``second'' coordination shell (against the Gaussian repulsion,
$1.5\leq r/\sigma \leq 3.5$) is present even at low density.

From a qualitative perspective, one might consider each core-softened
particle as effectively representing a cluster of water molecules
(e.g., a central water molecule and its four nearest
neighbors).\cite{Yan2007Correspondence-} In fact, Yan {\em et al.}
have recently presented evidence that a mapping of this sort has
quantitative merit when one compares, in appropriately reduced form,
the behaviors of a core-softened ramp model to TIP5P
water.\cite{Yan2007Correspondence-} When viewed from this perspective,
the formation of the first shell in the core-softened model at high
density qualitatively corresponds, in the molecular picture, to {\em
  additional} water molecules (5, 6, etc.)  penetrating the first
shell of an otherwise four-coordinated central water molecule.  Once this
physical relationship between the two models is recognized, the
similarities between their structural properties are easy to
understand.  To illustrate this, we carried out a structural analysis
of the core-softened model identical to that presented above for the
SPC/E model.

In particular, we examined the behavior of the RDF and $I_{s_2}$ [see
Fig.~\ref{fig:ramp_sim_s2fig}(b-g)] for the core-softened model as a
function of density along the $T^*=\kB T/\epsilon=0.3$ isotherm for
three different density regions: (1) the initial increase of
$\sTwo{\rho^*}$ at low densities [$\rho^*\leq 0.08$,
Figs.~\ref{fig:ramp_sim_s2fig}(b,c)], (2) the anomalous decrease of
$\sTwo{\rho^*}$ at intermediate densities [$0.08 \leq \rho^* \leq
0.175$, Figs.~\ref{fig:ramp_sim_s2fig}(d,e)], and (3) the increase of
$\sTwo{\rho^*}$ at high densities [$\rho^*>0.175$,
Figs.~\ref{fig:ramp_sim_s2fig}(f,g)].

As discussed above, the ``first'' shell of the RDF 
is not populated in this model at low density 
because the core-softened particles 
themselves loosely represent a central water and
its four nearest neighbors.  In this view, the initial compression of the 
core-softened fluid [region (1), $\rho^*\leq 0.08$] 
has an effect that is similar to that seen for SPC/E water.  
The modest increase in $-s_2/\kB$ that is observed 
is due to the increase in density and a minor enhancement of 
structuring in the second shell
($1.5\leq r/\sigma \leq 3.5$).    

Further compression of the core-softened model [region (2),
$0.08\leq\rho^*\leq 0.175$] leads to an anomalous decrease in
structural order [Fig.~\ref{fig:ramp_sim_s2fig}(a)].
Figs.~\ref{fig:ramp_sim_s2fig}(d) and (e) indicate that the
disordering is again due to a flattening and shifting inward of the second
shell.  Moreover, the ``first'' shell of the core-softened particles 
begins to emerge, which
schematically represents, in the approximate molecular view discussed above, 
that additional water molecules are effectively 
penetrating into the four-coordinated first shell. 

Similar to SPC/E water, it is known that there is a
strong correlation between excess entropy and self-diffusivity for the
core-softened model.\cite{Mittal2006Relationship-be} This information,
together with the results shown here, support the view that the
self-diffusivity anomaly of the core-softened model is also linked to
its density-dependent second-shell structure.

As expected, at higher density [region (3), $\rho^*\geq 0.175$],
compression leads to an increase in structural order due to
simple-liquid-like structuring of particles in the first coordination
shell [Figs.~\ref{fig:ramp_sim_s2fig}(f,g)].  In short, the
qualitative response to changes in density of the structural order and
its coordination-shell contributions for the core-softened model are
very similar to those of the more detailed SPC/E water model. This
finding is consistent with the recent demonstration that one can
approximately map the anomalies of TIP5P water onto those of a similar
two-scale ramp potential.\cite{Yan2007Correspondence-}

\begin{figure}[htb]
  \centering
  \includegraphics{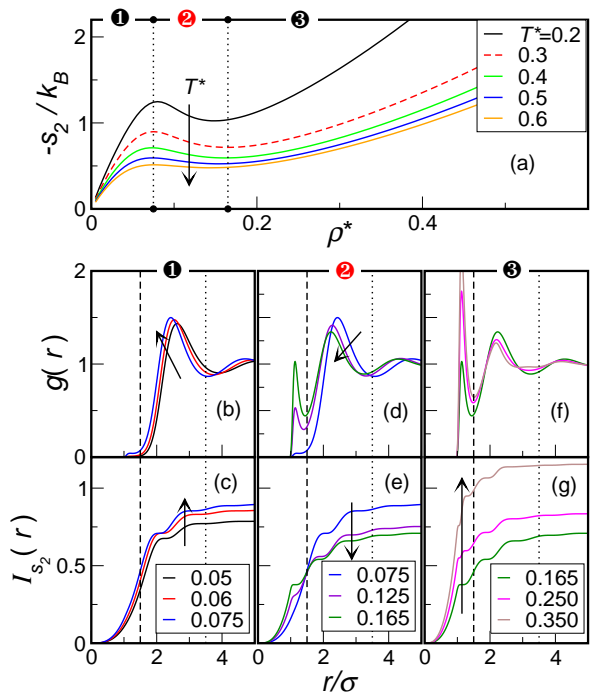}
  \caption{Structural data for the core-softened waterlike model from integral
    equation theory. (a) Structural order parameter $-s_2/\kB$ as a function
    of reduced density $\rho^*=\rho \sigma^3$ at the same values of
    $T^*=\kB T/\epsilon$ as in Fig.~\ref{fig:ramp_sim_s2fig}(a), where
    $\sigma$ is the particle diameter, and $\epsilon$ is the energy
    scale of the potential (see
    Eq.~\eqref{eq:core_softened_potential}). Arrow indicates direction of
    increasing $T^*$, and vertical dotted lines are at $\rho^*=0.075$
    and $\rho^*=0.165$, the approximate boundaries of the region of anomalous
    structural behavior. (Lower panel) Radial distribution function $g(r)$ and
    cumulative order integral $I_{s_2}(r)$ along the $T^*=0.3$ isotherm
    [red dashed curve in (a)] for three different density regions:
    (b,c) $\rho^* \leq 0.075$ [up to $-s_2(\rho^*)/\kB$ maximum], (d,e)
    $0.075 \leq \rho^* \leq 0.165$ [between maximum and minimum in
    $-s_2(\rho^*)/\kB$], (f,g) $\rho^*\geq 0.165$ [beyond minimum in
    $-s_2(\rho^*)/\kB$].  The regions are indicated by circled numbers
    along top of (a) and lower panel.  In lower panels, arrows indicate
    direction of increasing density; numbers in legends indicate
    values of $\rho^*$; vertical dashed line is at $r=1.5\sigma$ and
    vertical dotted line is at $r=3.5\sigma$, the approximate locations
    of the first and second minima in $g(r)$, respectively.}
  \label{fig:ramp_HNC_s2fig}
\end{figure}

As a final point, we show in Fig.~\ref{fig:ramp_HNC_s2fig} that the
integral equation theory of the core-softened model can qualitatively
predict all of the trends shown in Fig.~\ref{fig:ramp_sim_s2fig}.  The
ability of this approach to reproduce the structural features seen in
simulations, together with the quasi-universal connection between
structure and transport coefficients of
liquids,\cite{Rosenfeld1999A-quasi-univers,Rosenfeld1977Relation-betwee,Dzugutov1996A-univeral-scal}
suggests that integral equation theory might serve as a valuable
first-pass tool in assessing whether other model systems represent
good candidates for exhibiting static and dynamic anomalies.
However, if the intention is to ultimately use it as a quantitatively accurate 
predictive tool, then more comprehensive investigations of alternative
closure relationships, in the spirit of 
Ref.~\onlinecite{Oliveira2006Structural-anom}, will be necessary.   

\subsection{SRA fluids} 
\label{sec:struct-anom-short}
One of the key aspects of short-range attractive (SRA) fluids is that
their structurally anomalous behavior occurs as a function of the
reduced interparticle attractive strength $\epsilon/\kB T$ at constant
particle packing fraction $\Pc$, where $-\epsilon$ represents the well
depth of the interparticle attraction.  In most typical atomic or
molecular fluids, one finds that structural order ($-s_2/\kB$)
increases with $\epsilon/\kB T$.  SRA fluids are anomalous in that, at
sufficiently high values of $\Pc$, the opposite trend can be
observed;\cite{Krekelberg2007How-short-range,Mittal2006Quantitative-Li}
i.e., attractions counterintuitively decrease the amount of
structural order.

In this section, we briefly discuss how we used molecular simulation
and integral equation theory to gain insights into this trend.  We
accomplished this by exploring the various coordination-shell
contributions to $-s_2/\kB$ for the two model SRA fluids discussed in
Section~\ref{sra_models}.

\subsubsection{Colloid-polymer mixture}

\begin{figure}[htbp]
  \centering 
  \includegraphics{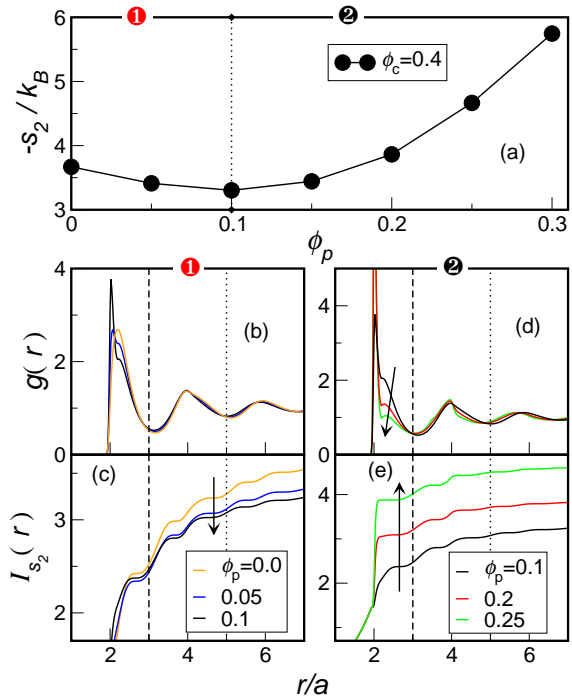}
  \caption{Structural data obtained from molecular dynamics simulations
    of the model colloid-polymer SRA fluid discussed in the text. (a)
    Structural order parameter $-s_2/\kB$ as a function of polymer
    volume fraction $\Pp$ (i.e., strength of colloid attractions) at
    colloid packing fraction $\Pc=0.4$. Vertical dotted line at
    $\Pp=0.1$, the location of the minimum in $\sTwo{\Pp}$. (Lower
    panel) Radial distribution function $g(r)$ and cumulative order
    integral $I_{s_2}(r)$ along the isochore $\Pc=0.4$ [black circles in
    (a)] for two polymer concentration ranges: (b,c) $\Pp\leq 0.1$
    [below minimum in $-s_2(\Pp)/\kB$], (d,e) $\Pp\geq 0.1$ [above
    $-s_2(\Pp)/\kB$ minimum]. The regions are indicated by circled
    numbers along top of (a) and lower panel.  In lower panels, arrows
    indicate direction of increasing $\Pp$; the parameter $a$
    indicates colloidal particle radius; vertical dashed line is at
    $r=3a$ and vertical dotted line is at $r=5a$, the approximate
    locations of the first and second minima in $g(r)$, respectively.}
  \label{fig:cfp_sim_s2fig}
\end{figure}

We begin by investigating the behavior of the model colloid-polymer
system\cite{Puertas2002Comparative-Sim,Puertas2003Simulation-stud} by
molecular simulation.  The effective colloid-colloid pair potential
for this model was presented earlier in Fig.~\ref{fig:CS_potential}(b)
for several values of polymer packing fraction $\Pp$.  Since the
reduced well-depth of this potential, $\epsilon/\kB T$, scales as
$\Pp$, we analyze structural order below as a function of the latter.

In particular, Fig.~\ref{fig:cfp_sim_s2fig}(a) illustrates how
$-s_2/\kB$ varies as a function of $\Pp$ at a particle packing
fraction of $\Pc=0.4$.  As expected for SRA fluids, $\sTwo{\Pp}$
exhibits a minimum at $\Pp\approx 0.1$.  In other words, this fluid
displays the structural anomaly of Eq.~\eqref{eq:anomaliesT} for
$\Pp\leq 0.1$. To understand the origins of this trend, we study the
RDF and the cumulative order integral $\Is2$ as a function of $\Pp$ in
two qualitatively different regions: (1) the anomalous decrease in
$\sTwo{\Pp}$ at low polymer concentrations (low interparticle
attractions), and (2) the ``normal'' increase of $\sTwo{\Pp}$ at
higher polymer concentrations (higher interparticle attractions).

What specific changes to coordination shell structure explain the
attraction induced disordering that occurs at small $\Pp$ [region (1),
$\Pp \leq 0.1$]?  First, note that strengthening interparticle
attractions considerably increases but narrows the first peak of the
RDF [Fig.~\ref{fig:cfp_sim_s2fig}(b)].  These two effects essentially
cancel so that the first-shell contributions to $\Is2$ are insensitive
to $\Pp$ over this range [see Fig.~\ref{fig:cfp_sim_s2fig}(c)].
However, attractions also slightly shift the higher coordination
shells of the RDF inward and diminish their overall correlation with
the central particle.  These latter modifications to the structure of
the second and higher coordination shells give rise to the anomalous
decrease in the structural order of this system.  They are also
consistent with behavior observed in the recent Krekelberg {\em et
  al.}\cite{Krekelberg2007How-short-range} simulations of the
square-well SRA fluid discussed in the Introduction.  A microscopic
interpretation of this trend is that SRA interactions drive weak
particle clustering at low $\Pp$ (explaining the sharpening and
narrowing of the first peak).  This clustering, in turn, opens up
channels of free volume in the fluid and disrupts the uniform
hard-sphere-like packing order in the second and higher coordination
shells.\cite{Krekelberg2006Model-for-the-f,Krekelberg2006Free-Volumes-an,Puertas2004Dynamical-heter,Sciortino2002One-liquid-two-}

Under conditions where the aforementioned structural anomaly occurs,
increases in $\Pp$ also increase the mobility of the
fluid.\cite{Krekelberg2007How-short-range} Very similar to the
waterlike fluids discussed above, it is known that $s_2$ and
self-diffusivity are strongly correlated for the model
colloid-particle mixture.\cite{Mittal2006Quantitative-Li} As a result,
the self-diffusvity anomaly appears to also derive from subtle
structuring effects in the second and higher coordination shells.

As one would expect, however, increasing $\Pp$ ultimately increases
structural order, if the interactions are sufficiently attractive [region
(2), $\Pp \geq 0.1$].  The attractions lead to the formation of
strongly bonded particle
clusters,\cite{Krekelberg2006Model-for-the-f,Krekelberg2006Free-Volumes-an,Puertas2004Dynamical-heter,Sciortino2002One-liquid-two-}
which is reflected by the increased height of the first peak of the
RDF [Fig.~\ref{fig:cfp_sim_s2fig}(d)] and the associated rise in the
first-shell contribution to $\Is2$ [Fig.~\ref{fig:cfp_sim_s2fig}(e)].

\begin{figure}[htb]
  \centering
  \includegraphics{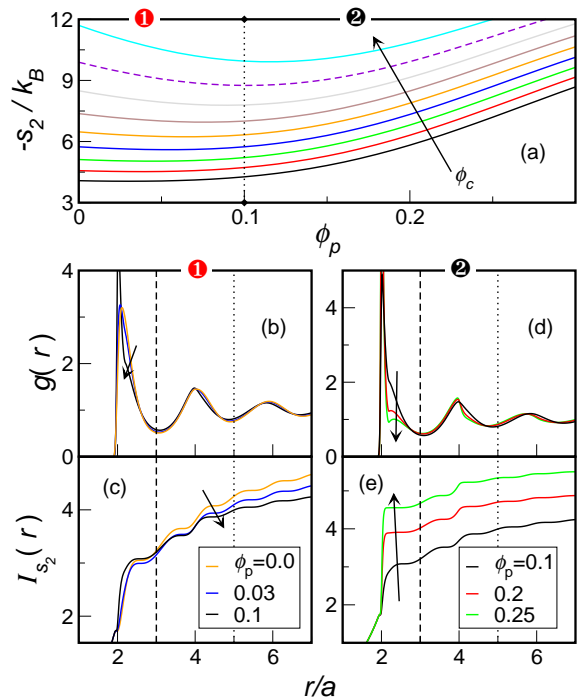}
  \caption{Structural data for the model colloid-polymer SRA fluid
    discussed in the text from integral equation theory.  (a)
    Structural order parameter $-s_2/\kB$ as a function of polymer
    volume fraction $\Pp$ at colloid packing fractions
    $\Pc=0.3,0.325,0.35,0.375,0.4,0.425,0.45,0.475$ and $0.5$.  Arrow
    indicates direction of increasing $\Pp$, and vertical dotted line
    is at $\Pp=0.1$, the approximate boundary of the region of
    anomalous structural behavior. (Lower panel) Radial distribution
    function $g(r)$ and cumulative order integral $I_{s_2}(r)$ along
    the isochore $\Pc=0.475$ [dashed violet curve in (a)] for two
    polymer concentration ranges: (b,c) $\Pp\leq 0.1$ [below minimum
    in $-s_2(\Pp)/\kB$], (d,e) $\Pp\geq 0.1$ [above $-s_2(\Pp)/\kB$
    minimum]. The regions are indicated by circled numbers along top
    of (a) and lower panel.  The parameter $a$ indicates colloid
    radius.  In lower panels, arrows indicate direction of increasing
    $\Pp$, vertical dashed line is at $r=3a$, and vertical dotted line
    is at $r=5a$, the approximate locations of the first and second
    minima in $g(r)$, respectively.}
  \label{fig:cfp_py_s2fig}
\end{figure}

\begin{figure}[htb]
  \vspace{0.2in}
  \centering
  \includegraphics{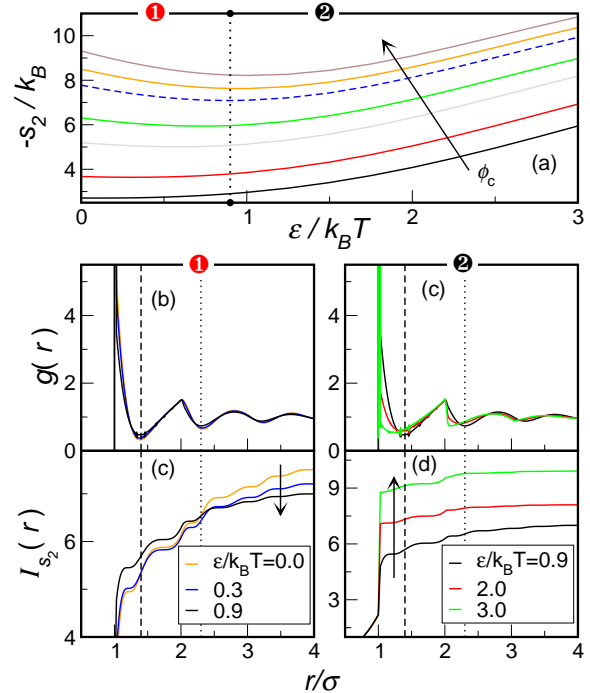}
  \caption{Structural data for the square-well fluid discussed in the
    text obtained from integral equation theory.  (a) Structural order
    parameter $-s_2/\kB$ as a function of reduced attractive strength
    $\epsilon/\kB T$ at particle packing fractions
    $\Pc=0.4,0.45,0.5,0.525,0.55,0.56$ and $0.57$.  Arrow indicates
    direction of increasing $\Pc$, and vertical dotted line is at
    $\ekt=0.9$, the approximate boundary of the region of anomalous
    structural behavior.  (Lower panel) Radial distribution function
    $g(r)$ and cumulative order integral $I_{s_2}(r)$ along the
    isochore $\Pc=0.55$ [dashed blue curve in (a)] for two attractive
    strength ranges: (b,c) $\ekt \leq 0.9$ [below minimum in
    $-s_2(\epsilon/\kB T)/\kB$], (d,e) $\ekt \geq 0.9$ [above
    $-s_2(\epsilon/\kB T)/\kB$ minimum]. The regions are indicated by
    circled numbers along top of (a) and lower panel. The parameter
    $\sigma$ indicates colloid diameter.  In lower panels, arrows
    indicate direction of increasing $\ekt$; vertical dashed line is
    at $r=1.4\sigma$ and vertical dotted line is at $r=2.3\sigma$,
    the approximate locations of the first and second minima in
    $g(r)$, respectively.}
  \label{fig:sw_py_s2fig}
\end{figure}

In closing, we test in Figs.~\ref{fig:cfp_py_s2fig} and
\ref{fig:sw_py_s2fig} whether integral equation theory is able to
qualitatively capture these attraction induced structural changes for
both model systems introduced in Section~\ref{sra_models}: the
colloid-polymer fluid and the square-well fluid, respectively.
Comparison of Fig.~\ref{fig:cfp_py_s2fig} with
Fig.~\ref{fig:cfp_sim_s2fig} and Fig.~\ref{fig:sw_py_s2fig} with
Figure~4 of Ref.~\onlinecite{Krekelberg2007How-short-range} demonstrate that
this is indeed the case. The success of these predictions strengthens
the case that integral equation theory will be a useful tool in
assessing whether future model systems of interest might display
structural anomalies.

\subsection{Conclusions}

Although the structural order of a fluid is usually enhanced by
isothermal compression or isochoric cooling, a few notable systems
show the opposite behaviors.  Specifically, increasing density can
disrupt the structure of waterlike fluids, while lowering temperature
(or strengthening of attractive interactions) can weaken the
correlations of fluids with short-range attractions.  The two-body
translational contribution to the excess entropy provides a
quantitative measure of these changes in structural order.  It is a
particularly insightful quantity to study because (i) its
contributions from the various coordination shells of the radial
distribution function can be readily determined, and (ii) it
correlates strongly with self-diffusivity, which allows it to provide
insights into the dynamic anomalies of these fluids.

Here, we have presented a comprehensive study, by both molecular
simulation and integral equation theory, of the coordination shell
contributions to the two-body excess entropy for several model
systems.  These models incorporate different levels of molecular
resolution, but all exhibit the aforementioned structural anomalies.
The results of this study support the emerging view that the
structural anomalies of these fluids can generally be attributed to
quantifiable changes in the second and higher coordination shells of
the radial distribution function.  They also demonstrate that integral
equation theory can serve as an effective first-pass tool for
assessing, based on the pair potential alone, whether new model
systems are good candidates for exhibiting static and dynamic
anomalies.

\begin{acknowledgments}
  We thank Anatol Malijevsky for providing an efficient code for
  numerical solution of the integral equation theories analyzed in
  this work.  W.P.K. acknowledges financial support of the National
  Science Foundation for a Graduate Research Fellowship.  T.M.T.
  acknowledges financial support of the National Science Foundation
  (CTS 0448721), the David and Lucile Packard Foundation, and the
  Alfred P.  Sloan Foundation. V.G. acknowledges financial support of
  the National Science Foundation (CTS-0347381) and the Robert A.
  Welch Foundation.  Computer simulations for this study were
  performed at the Texas Advanced Computing Center (TACC).
\end{acknowledgments}

\end{document}